\documentclass[twocolumn,aps,prl,showpacs]{revtex4}
\usepackage{amsmath}
\usepackage{amssymb}
\usepackage{color}
\usepackage[dvips]{graphicx}

\begin{document}

\title{Shaping the waveform of entangled photons}

\author{Alejandra Valencia}
\email{alejandra.valencia@icfo.es}
\author{Alessandro Cer{\` e}}
\altaffiliation{On the leave from Dip. di Fisica, Dipartimento di Fisica,
Universit\'a di Camerino, I-62032 Camerino, Italy}
\author{Xiaojuan Shi}
\author{Gabriel Molina-Terriza}
\altaffiliation{Also with ICREA-Institucio Catalana de Recerca i
Estudis Avancats, 08010 Barcelona, Spain}
\author{Juan P. Torres}
\altaffiliation{Also with Department of Signal Theory and
Communications, Universitat Politecnica de Catalunya, Barcelona,
Spain} 
\affiliation{ICFO-Institut de Ciencies Fotoniques,
Mediterranean Technlogy Park, Castelldefels, 08860 Barcelona,
Spain}







\date{\today}
\begin{abstract}
We demonstrate experimentally the tunable control of the joint
spectrum, i.e. waveform and degree of frequency correlations, of
paired photons generated in spontaneous parametric downconversion.
This control is mediated by the spatial shape of the pump beam in
a type-I noncollinear configuration. We discuss the applicability of this
technique to other sources of frequency entangled photons, such as
electromagnetically induced Raman transitions.
\end{abstract}


\maketitle

The quantum description of paired photons includes the spatial
shape, the polarization state and the joint spectrum. The later
contains all the information about bandwidth, type of frequency
correlations and waveform of the two-photon state. Quantum light
has been proved to be useful in many quantum information
applications and the most appropriate form of the joint spectrum
depends on the specific realization under consideration. For
example, uncorrelated pairs of photons can be used as a source of
heralded single photons with a high degree of quantum
purity~\cite{aichele1,urenlaser}; the tolerance against the
effects of mode mismatch in linear optical circuits can be
enhanced by using photons with appropriately tailored waveform
shape~\cite{rohde1}; the use of frequency-correlated or
anticorrelated photons allows erasing the distinguishing
information coming from the spectra when considering polarization
entanglement~\cite{grice1,poh}; some protocols for quantum
enhanced clock synchronization and positioning measurements rely
on the use of frequency anticorrelated~\cite{shihclock} or
correlated photons~\cite{lloyd1}. Moreover, the entanglement in
the frequency domain offers by itself a new physical resource
where to explore quantum physics in a high-dimensional Hilbert
space ~\cite{law1}. This requires the development of new
techniques for the control of the joint spectrum that will allow
the generation of multidimensional waveform alphabets.

The most widely used method for the generation of pairs of
entangled photons is spontaneous parametric down conversion
(SPDC). Notwithstanding, paired photons with the desired joint
spectrum may not be harvested directly at the output of the
downconverting crystal. The question that arises is how to control
independently different aspects of the joint spectrum of entangled
paired photons generated in SPDC; importantly, the sought-after
techniques should work for any frequency band of interest and any
nonlinear crystal.

Various methods have been proposed and developed to control the
type of frequency correlations and the bandwidth of downconverted
photons. Some of these methods rely on an appropriate selection of
the nonlinear crystal length and its dispersive
properties~\cite{grice1,kocuzu1}. Others are based on SPDC pumped
by pulses with angular dispersion \cite{torres1} or the design of
nonlinear crystal superlattices \cite{urenlattices}. Noncollinear
SPDC has also been propose as a way to tailor the waveform of the
downconverted
photons~\cite{walton1,uren2,carrasco1,lanco1,booth1,rossi1}.
Contrary to the case of collinear SPDC, where the transverse
spatial shape of the pump beam translate into specific features of
the spatial waveform of the two-photon state; in noncollinear
SPDC, the phase matching conditions inside the nonlinear crystal
mediates the mapping of spatial features of the pump beam into the
joint spectrum of the downconverted
photons~\cite{carrascospatspect}. This spatial-to-spectral mapping
allows to tune independently frequency correlations and the
waveform. In this letter, we demonstrate experimentally this
mapping and report experiments that demonstrate the feasibility of
using noncollinear SPDC as a tool to control the type of frequency
correlations using as tunable parameter the size of the pump beam
waist and the angle of emission of the downconverted photons. In
the past, measurements of the joint spectrum have been
reported~\cite{kim2,poh}. However, to the best of our knowledge
this is the first time that manipulations of the joint spectrum
have been demonstrated experimentally.


Let us consider noncollinear type-I SPDC in a nonlinear crystal of
length $L$ cut for non-critical phase matching. The spatiotemporal
quantum state of the two-photon pair can be written as
$|\psi\rangle=\int d\Omega_s d\vec{q}_{s} d\Omega_i d\vec{q}_{i}
\Phi \left( \Omega_s,\Omega_i, \vec{q}_{s},\vec{q}_{i} \right)
|\Omega_s,\Omega_i,\vec{q}_{s}, \vec{q}_{i} \rangle$, where $\Phi
\left(\Omega_s,\Omega_i,\vec{q}_{s},\vec{q}_{i}\right)$ is the
mode function or biphoton, which contains all the information
about the correlations and waveform properties of the two-photon
light. $\Omega_{j}=\omega_{j}-\omega_{j}^{0}$ are frequency
deviations from the central frequencies ($\omega_{j}^{0}$), and
$\vec{q}_{j}=\left(q_{jx},q_{jy} \right)$ are the transverse
wavevectors for the signal ($s$) and idler ($i$) photons.

In order to elucidate the frequency correlations and waveform of
the SPDC pairs, we consider the joint spectrum, $S \left(
\Omega_s,\Omega_i \right)=\left| \int d\vec{q}_{s} d\vec{q}_{i}
\Phi \left( \Omega_{s}, \Omega_{i},\vec{q}_{s}, \vec{q}_{i}
\right) U^{*} \left( \vec{q}_{s} \right) U^{*}\left( \vec{q}_{i}
\right) \right|^2$, where the function $U \left( \vec{q}_{j}
\right)$ describes the spatial mode in which the downconverted
photons are projected. For instance, gaussian modes when the
downconverted photons are collected with an imaging system
followed by single mode optical fibers \cite{collectionfiber1, collectionfiber2}. Projection into large area
modes is equivalent to projection into $
\vec{q}_{s}=\vec{q}_{i}\simeq0$, i.e., $U \left(\vec{q}\right)\propto \delta(\vec{q})$ .

\begin{figure}
\centering\includegraphics[scale=0.9,width=1\columnwidth]{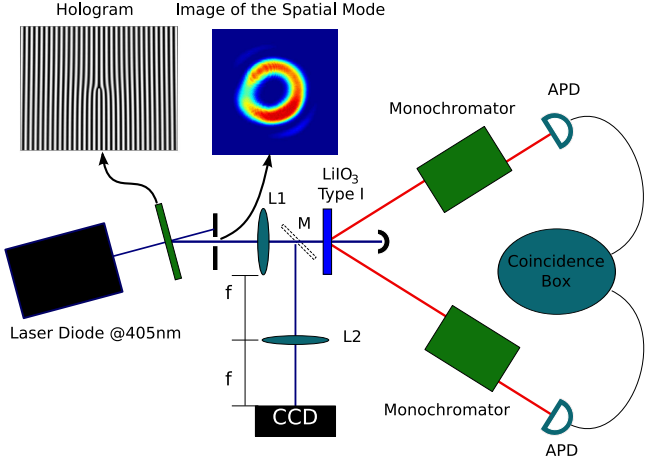}
\caption{Sketch of the experimental setup. The LG mode is
produced by a computer generated hologram. Photos of the hologram
and the diffracted beam are shown. The size of the beam impinging
into the crystal is controlled via L1. The downconverted beams are
collected into single mode optical fibers and sent through a pair
of computer controlled monochromators and APD detectors. Single
and coincidence counts are recorded with standard electronics. M is a flipping mirror used to switch from the frequency correlation measurement to the transverse momentum distribution measurement.}
\label{setup}
\end{figure}

The signal and idler photons travel
inside the crystal at an angle $\varphi_{s}=-\varphi_{i}=\varphi$ with
respect to the direction of propagation of the pump beam. The
mode function can then be written as \cite{torres2}
\begin{eqnarray}\label{biphoton}
 \Phi(\Omega_{s}, \Omega_{i},\vec{q}_{s}, \vec{q}_{i}) & =& E_{q}
\left( q_{xs}+q_{xi},\Delta_{0} \right) E_{\omega} \left( \Omega_s+\Omega_i \right) \nonumber \\
&  \times &\mathrm{sinc} \left( \frac{\Delta_{ k} L}{2} \right)
\exp \left\{ -i\frac{\Delta_{k} L}{2} \right\},
\end{eqnarray}
where $E_{q}$ and $E_{\omega}$ are the spatial shape of the pump
beam in the transverse wavevector domain and the pump pulse
frequency spectrum, respectively.
$\Delta_{0}=(q_{sy}+q_{iy})\cos\varphi-(k_{s}+k_{i})\sin\varphi$
accounts for the phase mismatching along the transverse direction, and
$\Delta_{k}=k_{p}-(k_{s}+k_{i})\cos\varphi-(q_{sy}+q_{iy})\sin\varphi$
for the phase matching conditions along the longitudinal direction. $k_{j}$
is the wavevector for signal, idler, and pump waves.
Eq.~(\ref{biphoton}) reveals that for the chosen configuration the
spatial properties of the pump, $E_{q}$, are mapped into the
spectral domain of the downconverted photons, $S \left(
\Omega_s,\Omega_i \right)$, through the dependence of $\Delta_0$
on the frequency.

To obtain further physical insight, we do a first order Taylor
expansion of $k_{j}(\omega_{j}^0+\Omega_j)$ around the central
frequencies, and assume large area collection modes
($\vec{q}_{s}=\vec{q}_i\simeq0$). The joint spectrum reduces to
\begin{eqnarray}\label{jointspectrumspatspect}
S \left( \Omega_s,\Omega_i \right) &=& \left| E_{q} \left(
0,N_{s}\sin\varphi \;\Omega_{-} \right) \right|^2
\left| E_{\omega}\left( \Omega_{+} \right) \right|^2 \nonumber \\
 & \times &\exp \left\{ -\frac{\left[ \alpha \left( N_{p}-N_{s}
\right) \cos \varphi \,L\right]^2}{4} \Omega_{+}^2 \right\},
\end{eqnarray}
where $N_{j}\equiv dk_{j}/d\omega_{j}$ is the corresponding
inverse group velocity, $\Omega_{+}\equiv\Omega_{s}+\Omega_{i}$
and $\Omega_{-}\equiv\Omega_{s}-\Omega_{i}$. We have approximated
the phase matching function, $sinc(\Delta_k L/2)$, by an
exponential function that has the same width at the $1/e^{2}$ of
the intensity: $sinc(b x)\simeq \exp[-(\alpha b)^{2}x^{2}]$, with
$\alpha=0.455$. Notice that the spatial to spectral mapping occurs
between the shape of the pump beam along the transverse direction and the
frequency shape in the $\Omega_{-}$ axis.

The experimental setup used to demonstrate this mapping is sketched
in Fig.~\ref{setup}. A $L=1$~mm long LiIO$_{3}$ crystal, cut for type-I
noncollinear degenerate SPDC, is pumped with a high power
laser diode (Nichia NDHV220APAE1) centered at $\lambda_p^0=405$~nm,
with a measured bandwidth of $\Delta \lambda_p \simeq 0.4$~nm.
The spatial mode of the laser beam is reduced to Gaussian by a set of
cylindrical lenses and a spatial filter. The degenerate down
converted photons centered at $\lambda_s^0=\lambda_i^0=810$~nm
are produced at an internal angle $\varphi=17.1^{\circ}$ (which
corresponds to noncritical phase matching) and are imaged into
two single mode fibers via two lenses ($f=11$~mm) placed at $54$~cm
from the output face of the nonlinear crystal. The output of
each fiber is sent through two monochromators (Jobin Yvon
MicroHR), and finally sent into single photon counting modules
(Perkin-Elmer SPCM-AQR-14-FC). Singles and coincidence counts for
the two detectors are recorded and the joint spectrum is
measured by scanning both monochromators.

\begin{figure}[t]
\centering\includegraphics[angle=-90,scale=1.2,width=1\columnwidth]{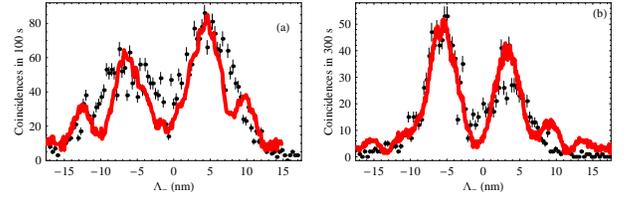}
\caption{Comparison of the measured spatial shape of the pump beam
in the transverse wavevector domain (solid line), and the measured
joint spectral intensity as a function of $\Lambda_{-}$ (dots).
(a) The pump beam shape is modified with a hologram and, (b) The
spatial shape is modified with a thin microscope slab.}\label{spatspect}
\end{figure}

In order to recognize the spatial-to-spectral mapping, we choose pump spatial modes whose transverse momentum distribution posses a clearly identifiable dip in the center.
The spatial profile of the pump beam is modify using two
different schemes. First, we use a hologram that generates, into
its first diffraction order, a vortex beam with topological charge
$m=2$. A picture of the hologram, and the corresponding shape of
the pump beam after the hologram, are shown in the inset of Fig.~\ref{setup}.
Alternatively, the pump beam is sent to a thin microscope slab
that introduced a phase shift over half of the beam.
The transverse momentum distribution of the pump, $E_{q} \left(
0,N_{s}\sin\varphi\; \Omega_{-} \right)$, can be measured by means of the $2 f$ system and a CCD camera.
The image from the camera relates to the frequency correlations according to $\Lambda_{-}= 4 \pi y/\left( f N_s \sin\varphi\; \omega_s^0 \right)$, where $y$ is the spatial coordinate in the transverse direction and $\Lambda_{-}\equiv
\lambda_s^0\, \Omega_{-}/\omega_s^0$.

Figure~\ref{spatspect} compares the measured spatial shape of the pump beam in the
transverse wavevector domain (solid line), $\left|E_{q} \left(
0,N_{s}\sin\varphi \Omega_{-} \right)\right|^{2}$, and the
measured joint spectrum shape as a function of $\Lambda_{-}$ (dots). Fig.~\ref{spatspect}a shows the results for the case where the pump
beam shape is modified with the hologram, and Fig.~\ref{spatspect}b when the
thin microscope slab is used. The matching of the solid lines and the points reveals that the spatial characteristics of the pump beam, $E_{q}
\left(0,N_{s}\sin \varphi \Omega_{-} \right)$, are mapped into the
joint spectrum of signal and idler photons,
$S(\Omega_s,\Omega_i)$. As a special feature, we can clearly see that the dip of the spatial pump profile translates into a dip in the joint spectrum at $\Lambda_{-}=0$, i.e. $\lambda_{s}=\lambda_{i}=810$~nm. 
Importantly, this technique make possible
to tailor the frequency waveform of the two-photon state
independently of the nature of the frequency correlations of the
photon pairs.

\begin{figure}[t]
\centering\includegraphics[angle=-90,width=0.9\columnwidth]{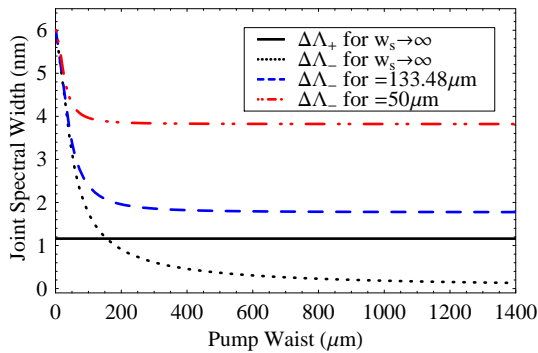}
\caption{Spectral widths $\Delta\Lambda_{-}$ and
$\Delta\Lambda_{+}$ as a function of the pump beam width ($W_0$).
$\Delta\Lambda_{-}$ is shown for three different values of the
spatial collection mode, $W_{s}=133.48$~$\mu$m, $W_{s}=50$~$\mu$m
and $W_s\rightarrow\infty$. Since $\Delta\Lambda_{+}$ does not
change significantly with $W_{s}$, only $\Delta\Lambda_{+}$ for
$W_s\rightarrow\infty$ is depicted. The intersection between
$\Delta\Lambda_{-}$ and $\Delta\Lambda_{+}$ for
$W_s\rightarrow\infty$ would allow complete tunable control of the
type of frequency correlations.} \label{curvetheory}
\end{figure}


The possibility of transferring the spatial characteristics of
the pump beam into the joint spectrum of the downconverted photons,
in principle allows the control of the type of frequency
correlations between signal and idler. Let us consider the case in
which the pump beam is characterized by a transverse momentum
profile $E(\vec{q}_{p})\propto \exp[-|\vec{q}_{p}|^{2}W_{0}^{2}/4]$
and a spectral distribution $E(\Omega_{p})\propto
\exp[-\Omega_{p}^{2}/(4B_{p}^{2})]$. $W_{0}$ and $B_{p}$ are the
beam waist and the bandwidth of the pump beam. Furthermore, signal and idler are
projected into the spatial mode
$U(\vec{q}_{j})\propto \exp[-|\vec{q}_{j}|^{2}W_{s}^{2}/4
]$. The joint
spectrum then reads

\begin{equation} \label{jointspectrumproject}
S \left( \Omega_s,\Omega_i \right)={\cal N} \exp \left\{
-\frac{\Omega_{+}^2}{2B_{+}^2} \right\} \exp \left\{
-\frac{\Omega_{-}^2}{2B_{-}^2} \right\}
\end{equation}
where
\begin{eqnarray}\label{bminus}
B_{-}=\left[ \frac{1}{2B_f^2}+\frac{\left( N_{s}\sin\varphi
W_{0}\right)^{2}}{1+2\left( W_{0}\cos\varphi/W_{s} \right)^{2}}
\right]^{-1/2},
\end{eqnarray}
\begin{eqnarray}\label{bplus}
B_{+}=\left[ \frac{1}{B_p^2}+\frac{1}{2B_f^2}+\frac{\left(\alpha L
\right)^2 \left( N_{p}-N_{s}\cos\varphi \right)^2}{1+2 \left(
\alpha \sin \varphi L/W_{s}\right)^2} \right]^{-1/2}
\end{eqnarray}
and $\cal N$ is a normalizing factor. We have also assumed the
presence of frequency filters of the form $H \left(
\Omega_{s,i}\right)\propto\exp \left[ -\Omega_{j}^{2} /\left(
4B_{f}^{2}\right) \right]$ in front of the detectors.

The ratio between the distribution widths $B_{+}$ and $B_{-}$ characterizes
the type of frequency correlation of the two-photon state:
anticorrelated photons are obtained for a pump beam width so
that $B_{-} \gg B_{+}$, while we get correlated photons
when $B_{-} \ll B_{+}$. If $B_{-}=B_{+}$, we have uncorrelated frequency photons.
From Eq.~(\ref{bminus}), the width $B_{-}$ mainly depends on the pump beam width and the
noncollinear angle. In order to compare with the
experimental data, we will work with the variables
$\Lambda_{-}$ and $\Lambda_{+}\equiv \lambda_{s}^0\Omega_{+}/\omega_{s}^0$ and
to which we associate the width (standard deviation) $\Delta\Lambda_{-}$ and
$\Delta\Lambda_{+}$, respectively. Fig.~\ref{curvetheory} shows
the dependence of $\Delta\Lambda_{-}$ on the
pump beam waist, $W_{0}$, for different values of $W_{s}$.
When for a value of $W_{s}$, the $\Delta\Lambda_{+}$ and $\Delta\Lambda_{-}$ curves cross
we have the possibility of genarating paired photons anticorrelated ($\Delta\Lambda_{+}<
\Delta\Lambda_{-}$), correlated
($\Delta\Lambda_{+}>\Delta\Lambda_{-}$), and even uncorrelated
($\Delta\Lambda_{+}=\Delta\Lambda_{-}$) in frequency by choosing an
appropriate values of the pump beam waist.

For the sake of clarity and due to the weak dependence of $\Delta \Lambda_{+}$ on $W_{s}$ for a given pump bandwidth,
we only plot $\Delta \Lambda_{+}$ for the ideal case, $W_{s}\rightarrow\infty$.
When comparing the curves for different values of
$W_{s}$, it is clear that the generation of
anticorrelated photons is not greatly affected by the collection
modes. Highly frequency anticorrelated photons are obtained
for a focused pump and the relationship $W_{s}/W_0$ is large so
that, effectively, one always projects into a large area mode and,
therefore, approaches the condition $\vec{q}_{s}=\vec{q}_{i}\simeq0$. On
the other hand, the size of $W_{s}$
sets a minimum value of the bandwidth of the pump for the generation of uncorrelated or highly
correlated frequency two-photon states.

\begin{figure}[t]
\includegraphics[angle=0,width=1\columnwidth]{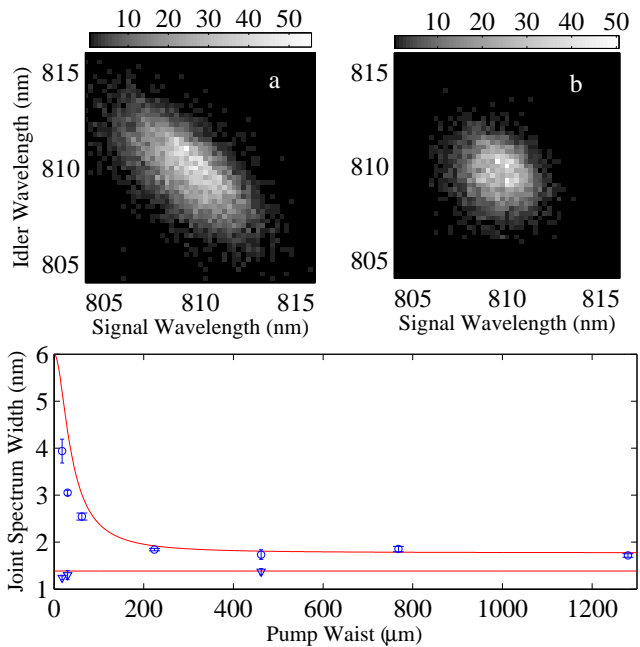}
\caption{Insets (a) and (b) corresponds to two different joint
spectra measured for different pump waists. (a) $W_0=30$~$\mu$m,
each point are coincidences measured in 50~s. (b)
$W_0=462$~$\mu$m, each point are coincidences measured in 200~s.
The curves in the lower part compares the theoretical prediction
for the variation of $\Delta \Lambda_{+}$  and $\Delta
\Lambda_{-}$ with $W_{0}$, with experimental values. In all cases,
the width of the Gaussian collection mode is $W_s=133$~$\mu$m.}
\label{Planes}
\end{figure}

We demonstrate experimentally the feasibility of frequency
correlation control exploiting the setup of Fig.~\ref{setup}. In
this case, we use a pump with a Gaussian profile and a telescope
to modify the pump beam waist. $W_{0}$ is measured by means of a
beam shaper (Coherent BM-7). For our collection scheme, the
imaging relation give us a collection mode waist,
$W_{s}^{exp}=133.48$~$\mu$m. The upper part of Fig.~\ref{Planes} shows
the joint spectra for two different values of $W_{0}$. These are
obtained by scanning the two monochromators and recording
coincidence counts. A two dimensional Gaussian fit is employed to
obtain the bandwidths along $\Lambda_{+}$ and
$\Lambda_{-}$ of the distribution. Part (a) of Fig. ~\ref{Planes}
corresponds to the case of frequency anticorrelated photons while
part (b) of Fig.~\ref{Planes} corresponds to a two-photon state
close to complete frequency uncorrelation. The values of the
bandwidth measured for Fig.~\ref{Planes}a are $\Delta
\Lambda_{+}=1.29$~nm and $\Delta \Lambda_{-}=3.05$~nm, and for
Fig.~\ref{Planes}b, $\Delta \Lambda_{+}=1.37$~nm and $\Delta
\Lambda_{-}=1.73$~nm.

In the lower part of Fig.~\ref{Planes}, we compare the measured
bandwidth for different pump beam waist, $W_{0}$, with the
theoretical prediction for our experimental parameters. From
Eq.~(\ref{bminus}), one obtains that for large values of $W_0$,
$\Delta \Lambda_{-} \rightarrow
\lambda_{s}^0/\omega_{s}^0\sqrt{2}/(N_{s}\tan \phi\; W_{s})$.
Under our experimental conditions, this asymptotic value is
$\Delta \Lambda_{-}^{\infty}=1.77$~nm. On the other hand, the
$\Delta \Lambda_{+}$ corresponding to the bandwidth of our pump
laser is $1.38$~nm. Therefore, the curves for $\Delta\Lambda_{+}$
and $\Delta\Lambda_{-}$ do not intersect, showing that our pump is
not broad enough for achieving complete frequency correlated
photon pairs with our $W_{s}$.


In conclusion, we have demonstrated experimentally the mapping of
spatial characteristics imprinted on the pump beam into the joint
spectrum of SPDC photons. Therefore, the generation of frequency
shaped waveforms by spatially shaping the profile of the pump
beam. Moreover, by making use of spatial light modulators, this
technique could be used for the implementation of engineered
multidimensional waveform alphabets with any type of frequency
correlations. 

We have extended this capability and we have shown
the feasibility of tunable control of the frequency correlations
of frequency-entangled two-photon states. The tuning parameter
that mediates the control of the joint spectrum and consequently,
the type of frequency correlations is the spatial beam waist of
the pump. By changing this tuning parameter we observed photons
with a highly reduced degree of frequency correlation. The role of
the spatial collecting mode and the bandwidth of the pump beam in
the generation of highly correlated photon pairs was also
explained.

The technique to control the waveform of entangled photon pairs
reported here can be of great interest for enhancing waveform
control of paired photons generated through two-photon Raman
transitions in electromagnetically induced transparency
schemes\cite{harris1}, where highly noncollinear configurations
are frequently used~\cite{harris2} and a rudimentary waveform
control is demonstrated.

\section*{Acknowledments}This work has been supported by
EC under the integrated project Qubit Applications (QAP, IST
directorate, Contract No. 015848), by the Government of Spain
(Consolider Ingenio 2010 QIOT CSD2006-00019 and FIS2004-03556),
and by the Generalitat de Catalunya.

\end{document}